\documentclass[10pt]{iopart}
\usepackage{iopams}  
\usepackage{graphicx}
\begin{document}

\title[INVERT nanostructure determination]{Nanostructure determination from the pair distribution function: A parametric study of the INVERT approach}

\author{Matthew J. Cliffe and Andrew L. Goodwin$^\ast$}

\address{Department of Chemistry, Inorganic Chemistry Laboratory, University of Oxford,
South Parks Road, Oxford OX1 3QR, U.K.}
\begin{abstract}
We present a detailed study of the mechanism by which the INVERT method [\emph{Phys.~Rev.~Lett.} {\bf 104}, 125501] guides structure refinement of disordered materials. We present a number of different possible implementations of the central algorithm and explore the question of algorithm weighting. Our analysis includes quantification of the relative contributions of variance and fit-to-data terms during structure refinement, which leads us to study the roles of density fluctuations and configurational jamming in the RMC fitting process. We present a parametric study of the pair distribution function solution space for C$_{60}$, \emph{a}-Si and \emph{a}-SiO$_2$, which serves to highlight the difficulties faced in developing a transferable weighting scheme.
\end{abstract}

\submitto{\JPCM}

\section{Introduction}
The absence of a generic methodology for determining the atomic-scale structure of disordered materials remains one of the key problems in contemporary structural science. Motivating the search for a solution is the desire to understand structure/property relationships in situations where disordered materials play a central role. Examples include biomineralisation processes \cite{Weiner_2005}, pharmaceutical polymorphism and stability \cite{Shalaev_2002,Bates_2006,Billinge_2010}, data storage in phase-change chalcogenides \cite{Sun_2006,Hegedus_2008,Matsunaga_2011}, and pressure- and temperature-induced amorphisation of oxide and metal--organic frameworks \cite{Keen_2007,Chapman_2009,Bennett_2010,Hu_2010,Bennett_2011,Bennett_2011c}. As these materials lack long-range periodic order, established crystallographic techniques cannot be used to produce structural models. Instead, techniques that are sensitive to the existence and nature of short-range structural correlations represent the only possibilities for experiment-driven structure determination.  It is for this reason that spectroscopic techniques such as NMR and EXAFS, together with the diffraction method of total scattering (often termed ``pair distribution function'', or PDF, measurements), have collectively played an important role in developing our understanding of disordered materials \cite{Brodsky_1978,Mullerwarmuth_1982,Shao_1990,Egami_2003}.

Of these various experimental techniques, PDF measurements arguably provide the most direct probe of local structural order \cite{Young_2011}. The PDF itself represents a histogram of interatomic separations, weighted by the relative concentration and scattering power of the different atom types present. So, even by using simple peak fitting methods it is possible to extract directly from the PDF the values of nearest- and next-nearest neighbour bond lengths and the corresponding coordination numbers. Coupled with chemical intuition, these values can often be used to infer longer-range and/or higher-order structural correlations, such as bond angles and network connectivity. But because the PDF is so straightforward to calculate from any given structural model, genuine structural refinement is now computationally tractable in a way that is not yet possible for many types of spectroscopic measurements (\emph{e.g.}\ NMR). Perhaps the most widely used refinement approach is that implemented in the software PDFGui \cite{Farrow_2007}. The general strategy employed by PDFGui bases refinement on the interatomic correlations present in a relevant periodic structure. The unit cell dimensions, atom coordinates and thermal displacement parameters are all refined against the PDF using a Rietveld-like algorithm. The absence of long-range order is then treated by some combination of restricting the fit to only the lowest-$r$  region and incorporating a damping term. This damping term contains useful information when it can be interpreted in terms of a characteristic coherence length-scale (which may itself have orientational dependence), reflecting \emph{e.g.}\ nanoparticle or domain size \cite{Farrow_2010}.

But both peak fitting and real-space Rietveld approaches have two serious shortcomings that in principle limit the scientific value of the structural information extracted from the PDF. The first is that neither actually produces a structural model in the form required for \emph{ab initio} electronic structure calculations or for molecular dynamics simulations. If the science of interest arises from understanding the relationship between structure and property in a material, this limitation can prove serious. Second, disordered materials can---and do---have different structures to their crystalline analogues \cite{Goodwin_2010}. Any approach that relies on using a periodic structure as its central reference inherently prohibits the exploration of disordered systems whose structures cannot be understood as nanocrystalline arrays.

The established alternative is to use an atomistic approach \cite{McGreevy_1988}. This involves calculating the PDF from a configuration of atoms that (i) is larger than the coherence length evident in the PDF, (ii) is subject to periodic boundary conditions, and (iii) is assembled using a composition and density appropriate for the material in question. The corresponding PDF is again straightforwardly calculated using the atomic coordinates but without need to apply any \emph{post hoc} damping correction. Refinement involves varying the atomic coordinates in this configuration until the best possible fit to data is achieved, often making use of a reverse Monte Carlo (RMC) algorithm in order to explore the large configurational space associated with perhaps thousands of positional parameters. In contrast to ``traditional'' structure refinements, there is no expectation that the PDF data are described by a unique set of atomic coordinates. Instead it is the set of general reproducible features (\emph{e.g.} topological connectivity, bond-length, bond-angle and torsional distribution functions) that is interpreted as the structure solution in this case. Ideally, the solution obtained should not depend on the starting configuration; nor should it be necessary to assume structural features (\emph{e.g.}\ coordination numbers and geometries) during refinement.

\begin{figure}
\begin{indented}
\item[]\includegraphics{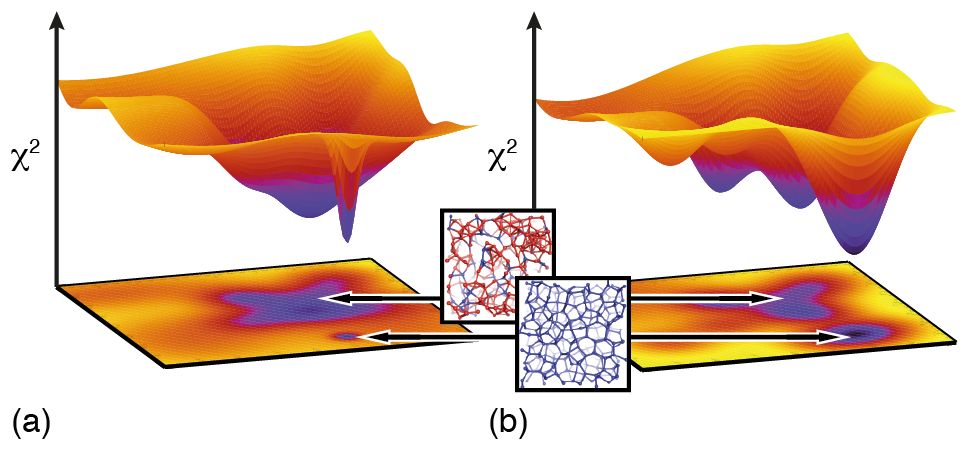}
\end{indented}
\caption{Schematic representation of the PDF refinement landscape. (a) Physically unreasonable (mostly red; incorrect local coordination) and sensible (blue; correct local coordination) configurations give rise to equally deep wells in configuration space, but the former are often more accessible. (b) The goal of nanostructure determination algorithms is to reshape solution space so as to increase the relative depth \emph{and} breadth of those minima associated with meaningful solutions.
 \label{fig1}}
\end{figure}

Arguably the most severe shortcoming of atomistic approaches such as RMC is the fact that meaningfully-different configurations can give rise to PDF fits of indistinguishable quality \cite{Gereben_1994}. Moreover, it is usually the case that the physically-sensible solutions are configurationally less accessible than are the vastly more numerous nonsensical solutions---\emph{so not only are incorrect solutions possible but they are in fact more likely}. In terms of the configurational landscape involved, we are describing a situation where the $\chi^2$ minima associated with meaningful structural solutions are few and steep, and where there are many equally-deep but broader minima associated with unphysical solutions [Fig.~\ref{fig1}(a)]. This observation frames the crucial challenge for nanostructure determination: namely, how does one redefine the energy landscape in a generic and transferable manner so as to ensure that the deepest minima always correspond to meaningful solutions and that these minima are configurationally accessible [Fig.~\ref{fig1}(b)]?

In a previous paper, we introduced the idea that experimental constraints on the number of unique environments might be used to guide structure refinement in a sensible way \cite{Cliffe_2010}. This approach---dubbed INVERT (= INVariant Environment Refinement Technique)---developed from a realisation that the single most obvious flaw in RMC configurations of canonical disordered systems such as amorphous silicon was their incorporation of an unphysical number of different coordination environments. Whereas the traditional remedy has been to enforce \emph{a priori} assumptions concerning the final structural model (\emph{e.g.}\ coordination numbers and geometries), we posited that the number and distribution of unique environments determined using spectroscopic measurements could be used as a constraint without needing to assume anything further about the nature of the environments themselves. For a handful of simple systems---C$_{60}$, S$_{12}$, \emph{a}-Si and \emph{a}-SiO$_2$---we demonstrated that a RMC+INVERT approach was markedly more effective than native RMC methods at arriving at sensible structure solutions.

In this paper, we explore in greater detail the mechanism by which INVERT actually guides nanostructure refinement, with a view to establishing how best the approach might be developed and applied in future studies. We begin by describing some of the different possible implementations of the central algorithm and then proceed to explore the question of algorithm weighting: namely, how does one balance the variance term with respect to the fit-to-data, and how is this balance affected by the presence of more than one type of chemical environment? In addressing this question, we quantify the relative contributions of variance and fit-to-data terms during structure refinement, and this leads us to study the roles of density fluctuations and configurational jamming in the RMC fitting process. We present a parametric study of PDF solution space for C$_{60}$, \emph{a}-Si and \emph{a}-SiO$_2$, which serves to highlight the difficulties faced in developing a transferable weighting scheme. Our paper concludes with a discussion of the major challenges and opportunities associated with developing further the INVERT approach in future studies.

\section{Implementation}
At its heart, the INVERT concept involves minimising the variance amongst the individual atomic PDFs for atoms in equivalent environments. Representing the individual PDFs by the term $g_i(r)$, one has for a single-environment system:
\begin{eqnarray}
\mathrm{Var}[g(r)]&=&\frac{1}{N}\sum_i\left[g_i(r)-\langle g(r)\rangle\right]^2\\
&=&\langle g_i(r)^2\rangle-\langle g(r)\rangle^2.
\end{eqnarray}
In the limit of zero variance, all of the $g_i(r)$ are equal to the same average $\langle g(r)\rangle$, and hence if this average PDF is to represent a good fit to the experimental PDF $G_{\mathrm{expt}}(r)$ then we essentially require that each $g_i(r)$ also match $G_{\mathrm{expt}}(r)$. This argument suggests a natural implementation of the INVERT approach in the case of a single environment, where the configuration quality is measured by the similarity of each atomic PDF to the experimental function:
\begin{equation}\label{inverteqn}
\chi^2_{\mathrm{INVERT}}=\sum_r\left\{\frac{1}{N}\sum_i\left[g_i(r)-G_{\mathrm{expt}}(r)\right]^2\right\}.
\end{equation}
We note at this point that the definition of the $g(r)$ and $G(r)$ is important, because the differences amongst $g_i(r)$ have a $r^{-1}$ dependence for glassy systems \cite{Levashov_2005}; in our work we use the definitions of Ref.~\cite{Keen_2001}, which enable direct comparison as suggested above. Recognising then that standard RMC algorithm involves minimising the function
\begin{equation}
\chi^2_{\mathrm{RMC}}=\sum_r\left[\langle g(r)\rangle-G_{\mathrm{expt}}(r)\right]^2,
\end{equation}
it is straightforward to show that Eq.~(\ref{inverteqn}) reduces to
\begin{equation}
\chi^2_{\mathrm{INVERT}}=\chi^2_{\mathrm{RMC}}+\sum_r\mathrm{Var}[g(r)].
\end{equation}

An extension to multiple environments also follows, where we assume initially that the experimental PDF can be decomposed into its constituent partial PDFs $G_{\mathrm{expt}}^{\alpha\beta}(r)$:
\begin{equation}
G_{\mathrm{expt}}(r)=\sum_{\alpha,\beta}c_\alpha c_\beta b_\alpha b_\beta G_{\mathrm{expt}}^{\alpha\beta}(r).
\end{equation}
Here $\alpha$ and $\beta$ index the different environments (or atom types), and the $c$ and $b$ are the corresponding relative concentrations and scattering strengths of those environments. The standard RMC penalty is
\begin{equation}
\chi^2_{\mathrm{RMC}}=\sum_r\left\{\sum_{\alpha,\beta}(c_\alpha c_\beta b_\alpha b_\beta)^2\left[\langle g^{\alpha\beta}(r)\rangle-G^{\alpha\beta}_{\mathrm{expt}}(r)\right]^2\right\},
\end{equation}
and hence
\begin{equation}
\chi^2_{\mathrm{INVERT}}=\chi^2_{\mathrm{RMC}}+\sum_r\left\{\sum_{\alpha,\beta}(c_\alpha c_\beta b_\alpha b_\beta)^2\mathrm{Var}[g^{\alpha\beta}(r)]\right\},
\end{equation}
as above.\label{cibi} What this suggests is that the $(c_\alpha c_\beta b_\alpha b_\beta)^2$ provide the natural weightings for the variance in each partial atomic PDF $g_i^{\alpha\beta}(r)$. We note that this weighting strategy can be applied even if the individual $G_{\mathrm{expt}}^{\alpha\beta}(r)$ are not experimentally separable.

\begin{figure}
\begin{indented}
\item[]\includegraphics{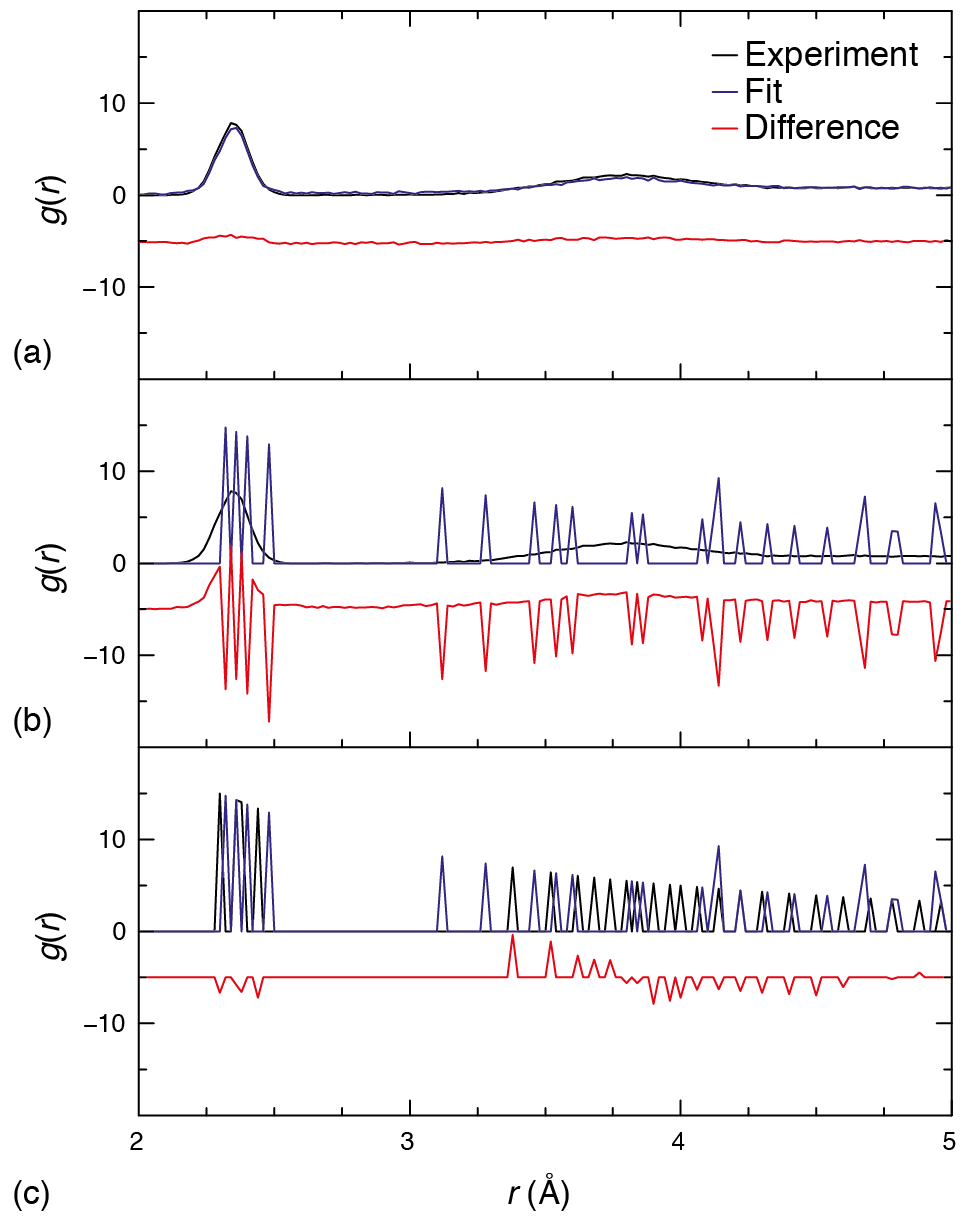}
\end{indented}
\caption{Three possible methods for fitting the PDF $G_{\mathrm{expt}}$ (black curve): (a) using the configurational average $\langle g(r) \rangle$, which is continuous in $r$. (b) using an individual atomic PDF $g_{i}(r)$, which gives rise to a discrete and unhelpful difference function. (c) using a distance list, which gives a discrete but meaningful difference function well suited to atomistic refinement. In each case the fit is shown in blue, and the difference function in red. \label{fig2}}
\end{figure}

In deriving the above equations, the implicit assumption is made that the $g_i(r)$ are directly comparable in a meaningful way. In a real material the functions $g_i(r)$ represent a time integral, such that each function could indeed resemble the time and configurational average $\langle g(r)\rangle$. The same is not true of course for static atomistic configurations such as used for RMC refinements: the calculated $g_i(r)$ consist of a series of delta functions that can be taken to represent an instantaneous $g_i(r,t)$. Ergodicity allows comparison of the configurational average $\langle g(r)\rangle$ with the experimental $G_{\mathrm{expt}}(r)$, but the same comparison is not meaningful for the $g_i(r,t)$ themselves [Fig.~\ref{fig2}].

One possible approach to resolving this problem is to convolve the $g_i(r,t)$ with a broadening function that represents the effect of thermal motion. For real materials, this motion is correlated and so the corresponding $r$-dependence would also need to be taken into account \cite{Jeong_1999,Jeong_2003}. Indeed, the sensitivity of the PDF to some directionality in the phonon dispersion also implies that additional orientation considerations may be required \cite{Goodwin_2004b,Goodwin_2005c}. The only unambiguous method of defining such a function is via direct calculation from a suitable lattice dynamical model \cite{Cope_2007}---a tractable, if unattractive, solution. However the broadening function may be defined, the key disadvantage of this approach is the computational cost of the convolution operation, which must be carried out for each atomic PDF at each step in the refinement.

An alternative approach---and the one we have mostly adopted---is to reformulate the PDF in terms of a distance list. This function is a discretised version of the PDF, which takes integral arguments: $d_{\mathrm{expt}}(n)$ can be defined as the distance $r_n$ for which
\begin{equation}
\int_0^{r_n}4\pi r^2\rho G_{\mathrm{expt}}(r)\,{\rm d}r=n,
\end{equation}
where $\rho$ is the number density. We note that a similar approach has been used in the Liga algorithm described elsewhere \cite{Juhas_2006,Juhas_2010}. The corresponding INVERT implementation can take one of two forms. Both involve calculating for each atom $i$ of type $\alpha$ the set of distances $d_i^{\alpha\beta}(n)$ to successive shells $n$ of neighbours of type $\beta$. The first possibility is to add to the standard $\chi^2_{\mathrm{RMC}}$ function an additional variance term of the form
\begin{equation}\label{networkinvert}
\chi^2_{\mathrm{Var}}=\sum_n\sum_{\alpha,\beta}\left\{\frac{A_{\alpha\beta}}{\langle d^{\alpha\beta}(n)\rangle^2}\sum_i\left[d_i^{\alpha\beta}(n)-\langle d^{\alpha\beta}(n)\rangle\right]^2\right\}.
\end{equation}
Here the $A_{\alpha\beta}$ represents a suitable weighting for the various partial PDFs (such as discussed above), and the additional $\langle d^{\alpha\beta}(n)\rangle^{-2}$ weighting is included to account for the fact that the number of neighbours grows as $r^2$ [Fig.~\ref{fig2}(c)]. This is the approach we have taken when fitting to \emph{a}-Si configurations. The second possibility is to fit the $d_i(n)$ directly to the $d_{\mathrm{expt}}(n)$ extracted from $G_{\mathrm{expt}}(r)$:
\begin{equation}\label{clusterinvert}
\chi^2_{\mathrm{INVERT}}=\sum_n\left\{\langle d_{\mathrm{expt}}(n)\rangle^{-2}\sum_i\left[d_i(n)-d_{\mathrm{expt}}(n)\right]^2\right\}.
\end{equation}
This is the general approach we have taken for small molecular clusters (C$_{60}$ and S$_{12}$), where the experimental $G_{\mathrm{expt}}(r)$ function is itself highly discrete and hence ineffective at guiding refinement.

A third possibility for comparing discrete individual PDFs to a time- or configurationally-averaged experimental PDF is to calculate the cumulative PDFs
\begin{eqnarray}
g_i^\prime(r)&=&\frac{1}{r}\int_0^rg_i(r^\prime)\,{\mathrm d}r^\prime\\
G^\prime_{\mathrm{expt}}(r)&=&\frac{1}{r}\int_0^rG_{\mathrm{expt}}(r^\prime)\,{\mathrm d}r^\prime.
\end{eqnarray}
We have found this to be particularly useful for disordered network systems where the configuration box has relatively few atoms: the fluctuations that impede direct fitting to $G_{\mathrm{expt}}(r)$ are now smoothed by the integral. In this paper, we apply this cumulative integral approach to the structure solution of \emph{a}-SiO$_2$.

\section{Fitting process}

While the formulations given in the preceding section suggest some natural weightings for the fit-to-data and variance $\chi^2$ contributions, we were interested to understand better the interplay between these two terms during the RMC fitting process. Here we are essentially asking two questions. First, to what extent does each term actually drive the refinement? And, second, do the two terms behave similarly for different systems: molecules \emph{vs} networks; systems with a single atom environments \emph{vs} those with multiple-atom environments?

\begin{figure}
\begin{indented}
\item[]\includegraphics{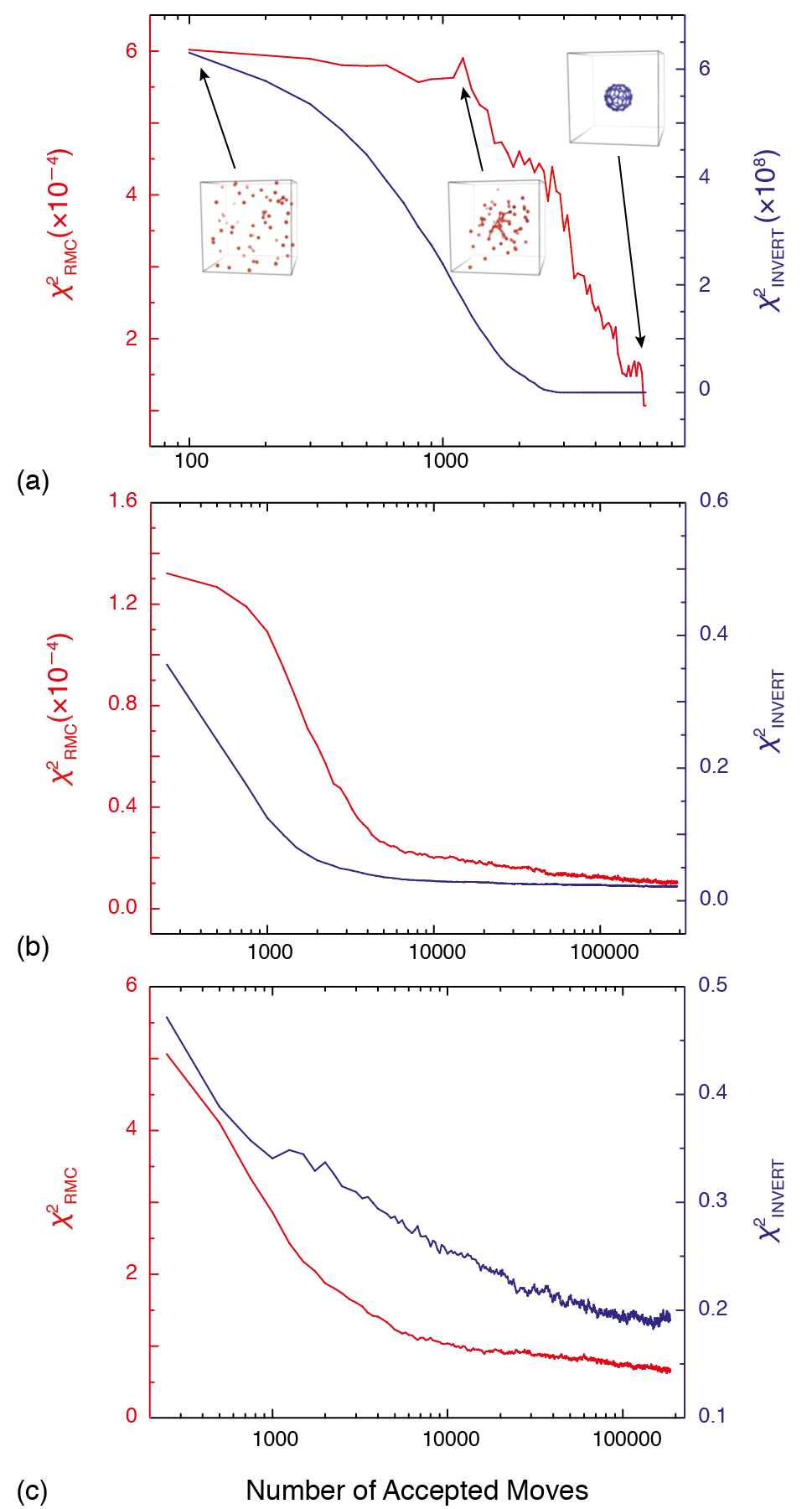}
\end{indented}
\caption{Evolution of fit-to-data and variance contributions to $\chi^2$ during RMC refinement of (a) C$_{60}$, (b) \emph{a}-Si and (c) \emph{a}-SiO$_2$.\label{fig3}}
\end{figure}

Considering first the ``nanostructured'' cluster C$_{60}$, our starting point is a large box containing 60 randomly-distributed C atoms. Because the system is molecular, we do not make use of periodic boundary conditions. The experimental PDF data used in our fit consist only of that portion that corresponds to intermolecular correlations (\emph{i.e.}\ $r<7$\,\AA) \cite{Juhas_2006}, and we apply the distance-list implementation of the INVERT variance function as described by Eq.~(\ref{clusterinvert}). The variation in $\chi^2$ components as a function of number of accepted moves is shown in Fig.~\ref{fig3}(a). What is clear is that the two components behave differently throughout the refinement process. The largest initial changes are to be found in the $\chi^2_{\mathrm{Var}}$ term; it is only once this has reached equilibrium that substantive reductions in $\chi^2_{\mathrm{RMC}}$ are at all observed. Examination of the configuration at different points in the refinement suggests that the initial decrease in $\chi^2_{\mathrm{Var}}$ corresponds to the formation of a single contiguous cluster of approximately spherical shape. The solution of the actual chemical structure of C$_{60}$ appears to be almost wholly driven by fitting to the PDF from this point onwards. Consequently, the failure of native RMC approaches to solve the structure of C$_{60}$ from PDF data \cite{Cliffe_2010} may largely reflect the configurational difficulty of clustering atoms from their initially-random positions in the RMC box.

The corresponding behaviour for the disordered network solids \emph{a}-Si and \emph{a}-SiO$_2$ is illustrated in Fig.~\ref{fig3}(b,c). In contrast to the trends observed for C$_{60}$ it seems that improvements to the fit-to-data and variance terms are more strongly coupled in these systems. If there is an obvious difference it is that the variance component in \emph{a}-SiO$_2$ shows a more gradual and sustained improvement throughout the refinement. This may reflect the more complex set of variance terms associated with a two-component system. In many ways the trend in $\chi^2_{\mathrm{expt}}$ is remarkably similar for the two refinements: a period of rapid initial improvement in fit-to-data (up to \emph{ca} 5000 moves) is followed by a sustained but noticeably more gradual improvement that persists throughout the remainder of the refinement.

\begin{figure}
\begin{indented}
\item[]\includegraphics{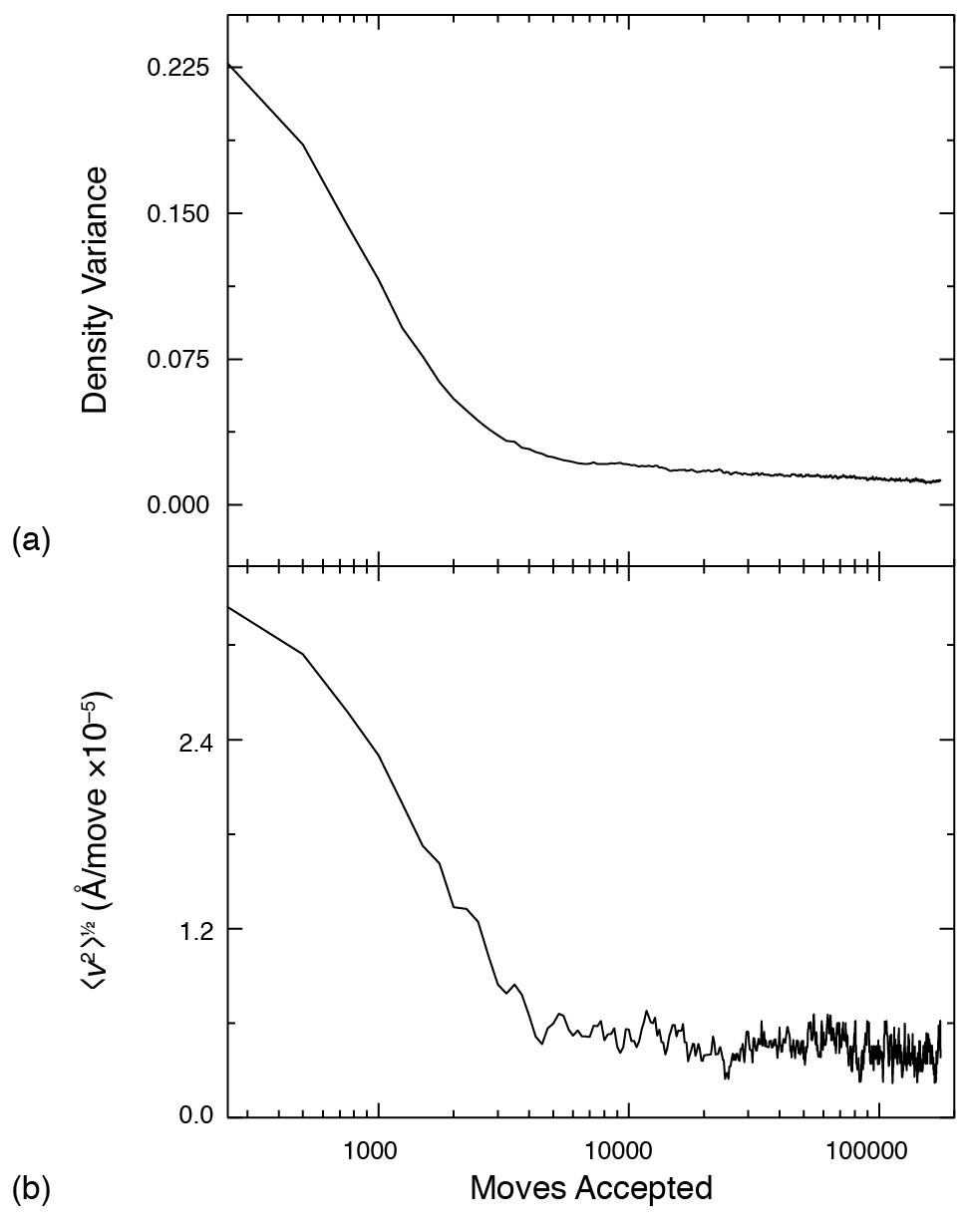}
\end{indented}
\caption{
Evolution of (a) density variance and (b) root mean squared displacement per accepted move (velocity, $v$) during RMC refinement of \emph{a}-Si.\label{fig4}}
\end{figure}

We proceed to demonstrate that for \emph{a}-Si these two refinement regimes reflect a period of initial density redistribution followed by subsequent small reorganisation of an essentially-jammed configuration. Our starting point is to consider the evolution in density variance throughout the refinement. This variance was calculated as follows. The RMC box was subdivided into a set of $50\times50\times50$ equally-sized voxels (\emph{ca} 0.4\,\AA\ each side). For each voxel, we counted the number of Si atoms contained within, and hence determined a voxel number density $\rho(i,j,k)$, where $i,j,k$ label the voxel in question. The density variance is then given by
\begin{equation}
\mathrm{Var}(\rho)=\sum_{i,j,k}[\rho(i,j,k)-\rho_0]^2,
\end{equation}
where $\rho_0$ is the total number density. The evolution of $\mathrm{Var}(\rho)$ as a function of accepted move is shown in Fig.~\ref{fig4}(a), from which it is clear that the first 5000-move phase of the refinement involves a monotonic reduction in density variance to a level that remains essentially unchanged thereafter. This initial configurational reorganisation is the only phase of the refinement for which large atom moves are accepted; the variation in mean ``velocity'' (= atom displacement per move) as a function of accepted move also reflects this result [Fig.~\ref{fig4}(b)].

\begin{figure}
\begin{indented}
\item[]\includegraphics{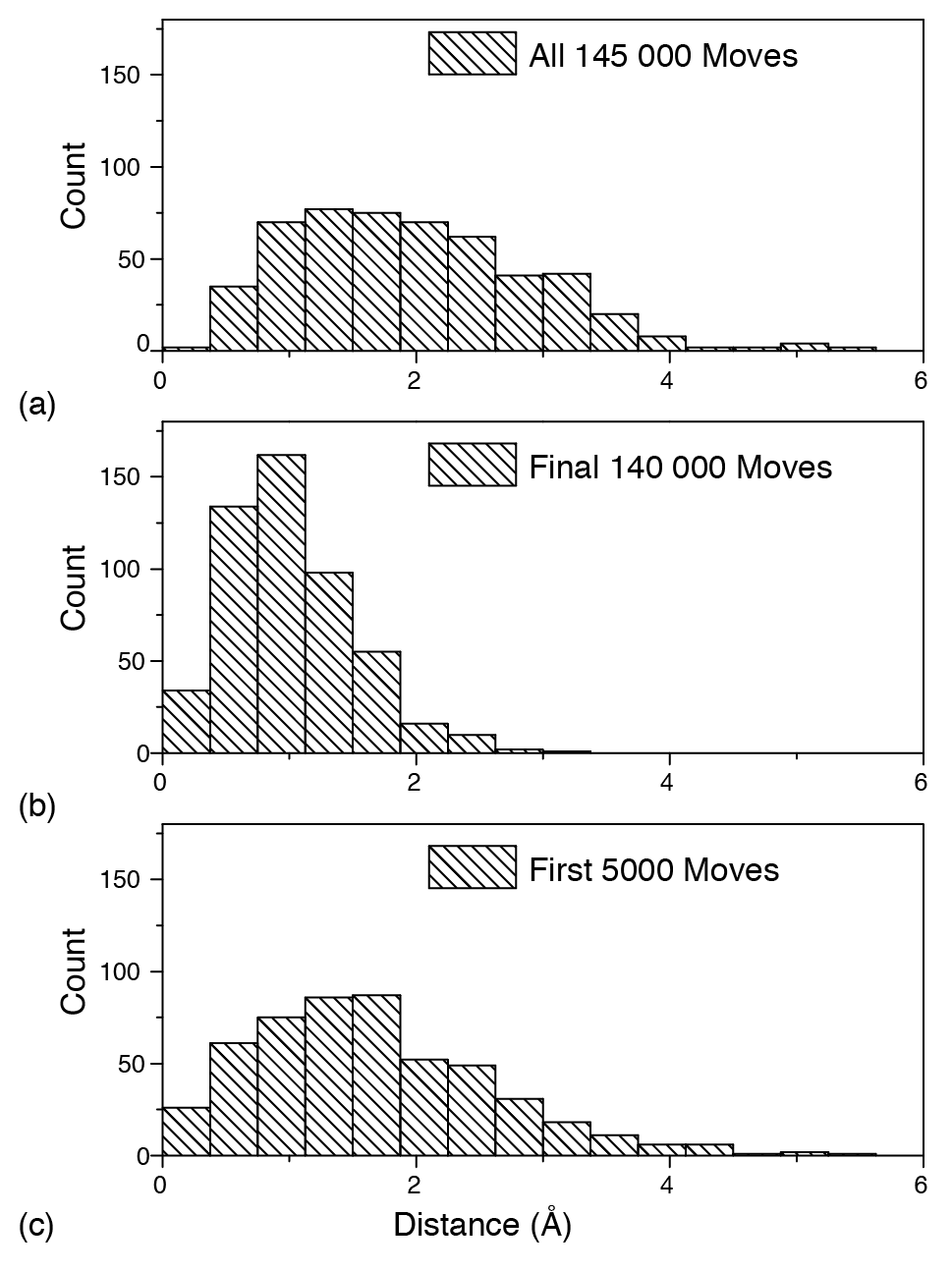}
\end{indented}
\caption{Trajectory length distributions for the (a) total, (b) final, and (c) initial RMC refinement regimes of \emph{a}-Si.\label{fig5}}
\end{figure}

That the configuration is essentially jammed after the first 5000 accepted moves is reflected in the histogram of trajectory lengths calculated for different regimes of the refinement [Fig.~\ref{fig5}]. Considering first the median displacement between initial and final atom positions, we find that most atoms have moved approximately 2\,\AA\ throughout the total course of the refinement (145\,000 accepted moves). The same histogram calculated for the final 140\,000 moves shows a markedly lower median value of less than 1\,\AA; whereas the corresponding histogram for the first 5000 moves is very similar to the total function, with a median displacement of 1.6\,\AA. Summing in quadrature yields a total displacement of 1.9\,\AA, indicating that there is very little correlation between moves in the two phases. So the displacements after 5000 moves reflect small structural reorganisations without any significant changes to network topology. Our final comment with respect to this refinement concerns the number of accepted moves themselves. We find that the logarithm of the number of accepted moves is not a linear function of the logarithm of the number of generated moves, but rather is linear only over two separate regimes that correspond respectively to fewer than, and more than, 5000 accepted moves.

So what are the implications of the existence of these two regimes of RMC refinement? The obvious concern is that the system does not have the requisite flexibility to explore configurational space adequately. This means that the atom positions in the final configuration are likely to depend strongly on the fluctuations present in the initial starting configuration. To test this hypothesis, we set up a refinement for \emph{a}-Si for which the initial configuration was based on the structure of crystalline silicon itself. We found that the system was unable to eradicate the bias of its initial periodicity, giving a fit-to-data that was approximately 10\% worse than for refinements set up using random initial configurations. The final variance term is similar for both types of refinement (of course $\chi^2_{\mathrm{Var}}$ is initially zero-valued for the crystalline starting configuration), so it is clear that refinement is actually jammed because the total $\chi^2$ value is higher than it might otherwise be. It might be hoped that the appropriate choice of move strategy might be able to reduce the propensity of the refinement to jam. We therefore tried a variety of different maximum move sizes, and also experimented with including a small chance of large moves. We found that this made little difference to the final configurations. Perhaps the most interesting conclusion to be drawn here is that the behaviour of the variance term in the early refinement region---where it is acting to increase homogeneity---is advantageous in the case of a molecular system, but may in fact contribute to the jamming problem for networks.

\section{Solution spaces}

We return now to the problem of optimising the relative weights of fit-to-data and and local variance $\chi^2$ terms. In our earlier study \cite{Cliffe_2010}, we used an \emph{ad hoc} method to explore the parameter space created by the weightings of these terms. While it was the case that we obtained satisfactory refinements, we did not know whether the weightings we had chosen were the optimum values, or indeed how sensitive the solution space was to variation in those parameters. Sensitivity becomes an especially important factor in the instance that there is no single transferable weighting scheme. So the approach we take here is to explore the refinement ``success'' for C$_{60}$, \emph{a}-Si and \emph{a}-SiO$_2$ as a function of data and variance weightings. Our interest is in the quality of the configuration, rather than the magnitude of $\chi^2$ and so we use as our metric the number of atoms with the correct coordination number. In the case of C$_{60}$, configurations with 60 (or nearly 60) threefold-coordinated C atoms universally correspond to good structural models with the expected truncated icosahedral shape. For the network solids, coordination number remains a useful metric although we note that in itself it is blind to unphysical features such as the existence of triangular ring structures.

\begin{figure}
\includegraphics{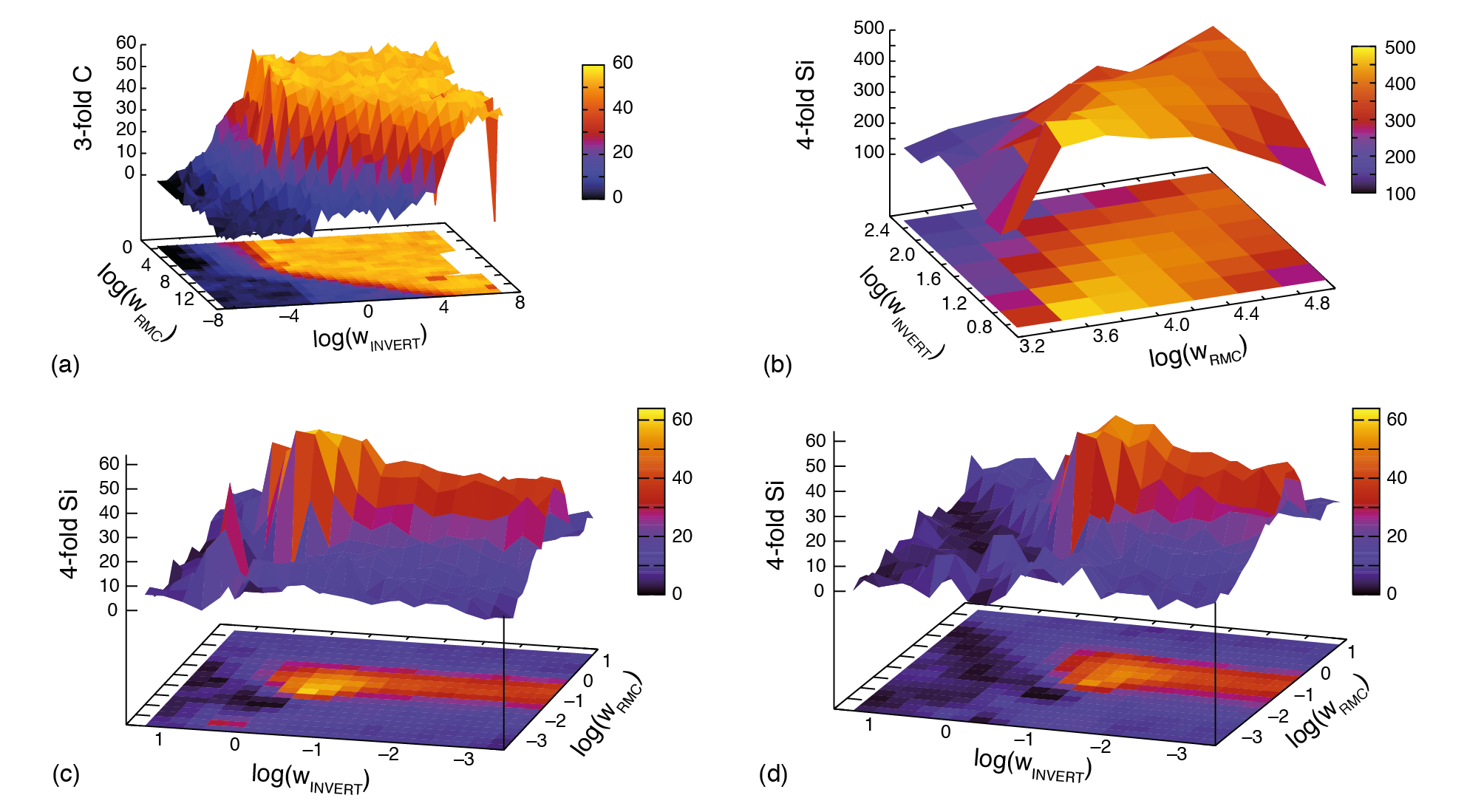}
\caption{\label{fig6}. Representations of the solution spaces for refinements of (a) C$_{60}$, (b) \emph{a}-Si, and (c,d) \emph{a}-SiO$_2$. Panel (a) shows the number of triply-coordinated C atoms in RMC configurations for C$_{60}$ after 10$^6$ proposed moves; panel (b) gives the number of fourfold coordinated Si atoms in configurations of \emph{a}-Si (total 512 atoms); panels (c) and (d) give the number of fourfold coordinated Si atoms in configurations of \emph{a}-SiO$_2$ (total 64 Si atoms and 128 O atoms) with pariwise variance terms weighted equally in (c) but by the relevant $b_ib_jc_ic_j$ terms in (d).}
\end{figure}

Considering first the solution space we observe for C$_{60}$, we find a remarkably well-defined plateau that extends across some 14 orders of magnitude of different weighting values [Fig.~\ref{fig6}(a)]. Convergence appears to be predicated primarily on the INVERT constraint: in particular, in its absence there appears to be no prospect of structure solution. The ability of the INVERT term to drive structure solution by itself is surely a result of the specific implementation we have used in this case: the $\chi^2$ formalism of Eq.~(\ref{clusterinvert}) includes an implicit fit-to-data term. Indeed the only effect of weighting the data contribution more strongly is to reduce the likelihood of obtaining the correct solution. So it appears that in this instance the natural weighting of INVERT and fit-to-data is precisely that needed to obtain a reproducible and accurate structure solution.

As suggested by their different refinement behaviour, the parameterisation plots we obtain for \emph{a}-Si and \emph{a}-SiO$_2$ reveal very different sensitivities to that observed for C$_{60}$ [Fig.~\ref{fig6}(b,c)]. For both of the network solids, there appears to exist a well-defined optimum set of weights that is not obviously related between the two systems. Moreover, the region of maximum coordination number exists only over a relatively narrow range of weightings, and---especially in the case of \emph{a}-SiO$_2$---it can also lie adjacent to other regions of very poor solution quality (note the rapid variation in final coordination number observed for different values of $\log(w_{\mathrm{PDF}})$ at $\log(w_{\mathrm{INVERT}})\sim-1$). Because we are making use of the INVERT implementation outlined in Eq.~(\ref{networkinvert}), which does not contain any implicit fit-to-data component, we find for these network structures that sufficiently high weightings of $\chi^2_{\mathrm{INVERT}}$ lead to unphysical structures.

Because \emph{a}-SiO$_2$ contains atoms that exist in two different local environments (Si and O), we used this system to establish the extent to which its solution space is affected by different weighting schemes for the various pairwise variance terms. We have explored two schemes in particular: the first makes use of identical weights for each of the four variance terms; the second applies a weighting based on the $c_ic_jb_ib_j$  terms outlined on page~\pageref{cibi}. The corresponding solution space plots are given in Fig.~\ref{fig6}(c,d), where it is clear that their overall form is essentially unchanged by the variation in weighting scheme. Intriguingly, the variation with fit-to-data weighting is almost identical in the two approaches; instead the most obvious difference is the location of the global maximum along the INVERT weighting axis. The difference in optimal $\log(w_{\mathrm{INVERT}})$ values of approximately 0.75 is much smaller than the logarithm of the square of the $c_ic_jb_ib_j$ terms used (\emph{ca} 3). Consequently the effective INVERT weighting at the maximum in Fig.~\ref{fig6}(d) is larger than that at the maximum in Fig.~\ref{fig6}(c). This suggests that the crucial pairwise variance is relatively weakly weighted amongst the $c_ic_jb_ib_j$. The smallest of these corresponds to the Si/Si pairs, so the implication here is that it is the arrangement of Si atoms around each other that is critical in defining the structure of \emph{a}-SiO$_2$.

Having established the effects of the various weightings employed during an RMC+INVERT refinement, we subsequently explored the role of configuration box size. To this end, we compared the outcomes for \emph{a}-Si refinements based on configurations containing 64 and 512 atoms. Variation in box size has a number of consequences for the refinement strategy. Clearly the speed of convergence is much faster for small boxes. But there are more subtle differences as well. The quantity of data used in the fit also differs, for example, because $r_{\mathrm{max}}\propto N^{1/3}$. It is not necessarily straightforward to adjust weightings to compensate fully for this change as variation in $N$ also affects the discreteness of the calculated $G(r)$, which in turn affects how closely the data can be fitted. Nevertheless we find that the optimal ratio of data and INVERT weighting for both 64 and 512 atom systems is roughly similar (the logarithmic difference between the two weightings is 2.55 for the former and 3.3 for the latter). Consequently the shape of the solution space appears to be largely unaffected by configuration box size.

\begin{figure}
\includegraphics{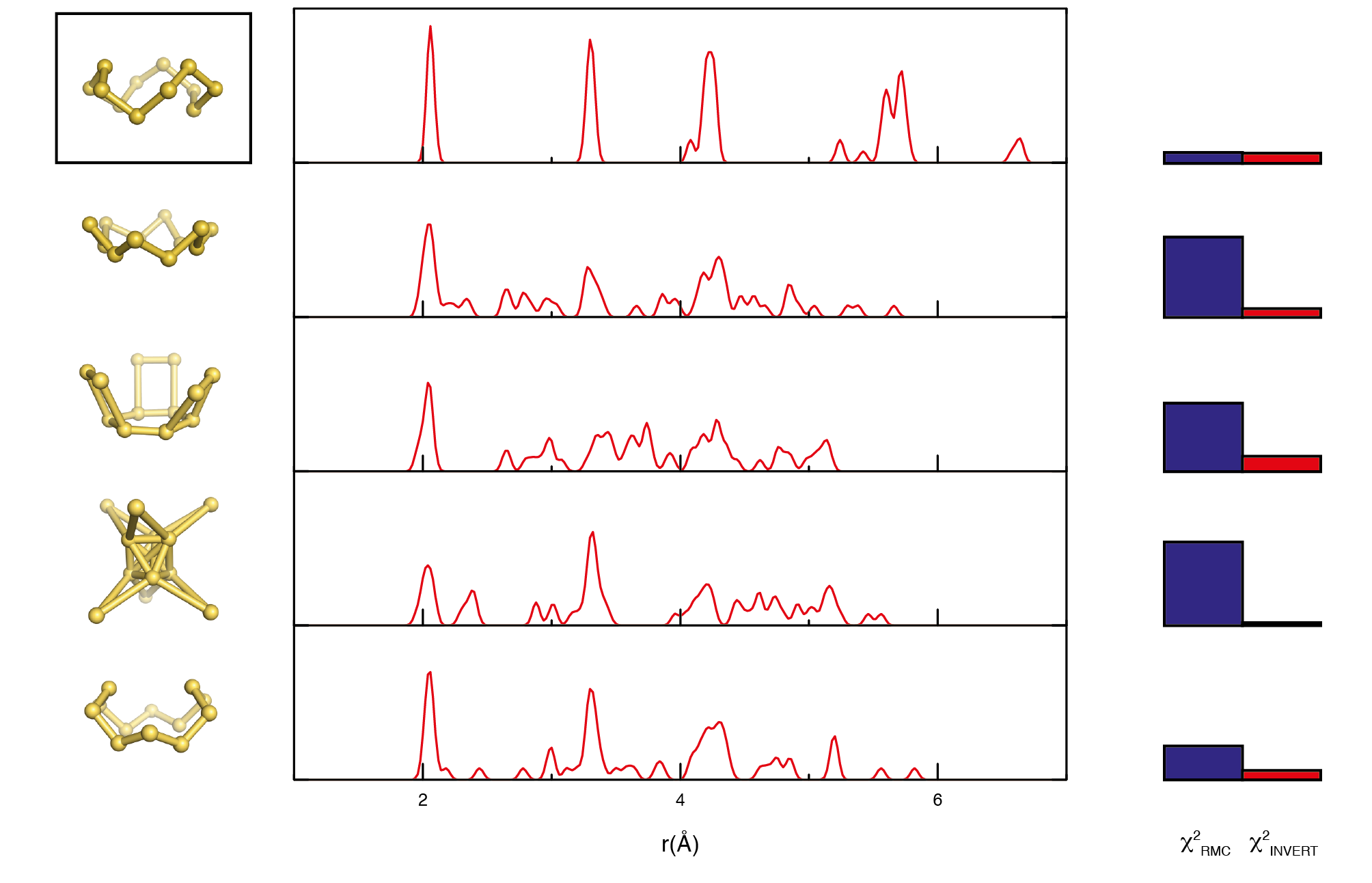}
\caption{Illustrations of molecular geometries and PDFs associated with local minima in the INVERT-weighted configuration space for S$_{12}$. The blue and red vertical bars represent the corresponding relative magnitudes of, respectively, the $\chi^2_{\mathrm{RMC}}$ and $\chi^2_{\mathrm{INVERT}}$ terms. The global minimum occurs for the correct geometry, which is highlighted at the top.\label{fig7}}
\end{figure}

We conclude this section by flagging the existence of a range of interesting near-solutions in the case of S$_{12}$. We first investigated this molecule as an example of a small nanostructured system with a single type of atom in two environments: its ring structure contains equatorial and axial S atoms that have similar nearest-neighbour distances but slightly-different next-nearest-neighbour separations. While the global minimum in configurational space does indeed correspond to the correct structure, the small molecule size results in a number of distinct local minima with intriguing structures (a number of which are shown in Fig.~\ref{fig7}). Clearly the chemistry of these candidate structures would be very different! But our purpose in highlighting their existence here is to provide a caveat for small-molecule structure solution using PDF data that the solution space may well be densely populated by diverse candidates with surprisingly similar PDFs but very different three-dimensional structures.

\section{Outlook and Conclusions}

Arguably one of the most appealing features of the INVERT approach when it was first proposed was the apparent transferability of the same simple idea across the very different ``nanostructure'' problems of molecular phases and disordered networks alike. Yet perhaps the clearest results of this extended study have been the demonstration that INVERT can work very differently in these different systems, and also that the balance between favouring local structural invariance on the one hand, and producing the best-quality fit to data on the other hand, is not only fine but is also system-dependent. Furthermore, we have shown that RMC refinement becomes effectively jammed after a relatively small number of accepted moves, such that configurational space is not yet effectively explored. This has the important consequence that structure solutions may depend on features of the initial configuration; and so we are probably not yet in a position to claim that nanostructure ``solution'' is possible at all using PDF methods.

But we have also tried to show that there are many feasible implementations of the INVERT approach---even within the limited definition of pairwise correlations we have explored here. It is obvious that the approach is already capable of helping guide refinement (C$_{60}$ being a particularly successful example), and one obvious avenue for further research is to determine more robustly which of the various $\chi^2_{\mathrm{Var}}$ formalisms is most effective in this respect. It may also prove beneficial to re-evaluate the RMC move strategy: in particular, is it possible to sample configurational space more effectively than with the standard algorithm of small individual moves, thus avoiding the difficulties associated with jammed refinements? And, finally, we note the suggestion we have made elsewhere \cite{Cliffe_2013} that consideration of higher-order correlation functions and local symmetry in an INVERT-type refinement may prove useful avenues for further improving the method; this is an area of research we are actively pursuing. 

\ack{The authors are grateful to the EPSRC and the ERC for financial support under grants EP/G004528/2 and 279705, respectively.}

\section*{References}

\bibliography{jpcm_2013_invert}
\bibliographystyle{unsrt}

\end{document}